\documentclass[a4paper]{article}
\usepackage{INTERSPEECH2021}
\usepackage{xcolor}
\usepackage{url}
\usepackage{color, colortbl}
\usepackage{graphicx}
\usepackage{todonotes}
\usepackage{caption}
\usepackage{subcaption}
\usepackage{tipa}

\let\OLDthebibliography\thebibliography
\renewcommand\thebibliography[1]{
  \OLDthebibliography{#1}
  \setlength{\parskip}{0pt}
  \setlength{\itemsep}{0pt plus 0.3ex}
}

\usepackage{flushend}

\newenvironment{alphafootnotes}
  {\par\edef\savedfootnotenumber{\number\value{footnote}}
   
   \setcounter{footnote}{0}}
  {\par\setcounter{footnote}{\savedfootnotenumber}}

\makeatletter
\def\blfootnote{\gdef\@thefnmark{}\@footnotetext}
\makeatother

\makeatletter
\def\bstctlcite{\@ifnextchar[{\@bstctlcite}{\@bstctlcite[@auxout]}}
\def\@bstctlcite[#1]#2{\@bsphack
  \@for\@citeb:=#2\do{%
    \edef\@citeb{\expandafter\@firstofone\@citeb}%
    \if@filesw\immediate\write\csname #1\endcsname{\string\citation{\@citeb}}\fi}%
  \@esphack}
\makeatother

\title{Investigating the Utility of Multimodal Conversational Technology and Audiovisual Analytic Measures for the Assessment and Monitoring of Amyotrophic Lateral Sclerosis at Scale}
\name{Michael Neumann$^{\dagger}$, Oliver Roesler$^{\dagger}$, Jackson Liscombe$^{\dagger}$, Hardik Kothare$^{\dagger\ddagger}$, David Suendermann-Oeft$^{\dagger}$, David Pautler$^{\dagger}$, Indu Navar$^a$, Aria Anvar$^a$, Jochen Kumm$^b$, Raquel Norel$^c$, Ernest Fraenkel$^d$, Alexander V. Sherman$^{e,f}$, James D. Berry$^e$, Gary L. Pattee$^{g}$, Jun Wang$^{h}$, Jordan R. Green$^{e,f}$ and Vikram Ramanarayanan$^{\dagger\ddagger}$}

\email{vikram.ramanarayanan@modality.ai}

\begin{document}

\bstctlcite{IEEEexample:BSTcontrol}

\maketitle

\blfootnote{$^{\dagger}$Modality.ai, Inc., San Francisco, United States}
\blfootnote{$^{\ddagger}$University of California, San Francisco}

\begin{alphafootnotes}
\footnotetext[1]{EverythingALS, Peter Cohen Foundation}
\footnotetext[2]{Pr3vent, Inc.}
\footnotetext[3]{IBM Thomas J. Watson Research Center}
\footnotetext[4]{Massachusetts Institute of Technology}
\footnotetext[5]{MGH Institute Health Professions}
\footnotetext[6]{Harvard University}
\footnotetext[7]{University of Nebraska}
\footnotetext[8]{University of Texas at Austin}
\end{alphafootnotes}

\begin{abstract}
 We propose a cloud-based multimodal dialog platform for the remote assessment and monitoring of Amyotrophic Lateral Sclerosis (ALS) at scale. This paper presents our vision, technology setup, and an initial investigation of the efficacy of the various acoustic and visual speech metrics automatically extracted by the platform. 82 healthy controls and 54 people with ALS (pALS) were instructed to interact with the platform and completed a battery of speaking tasks designed to probe the acoustic, articulatory, phonatory, and respiratory aspects of their speech. 
 We find that multiple acoustic (rate, duration, voicing) and visual (higher order statistics of the jaw and lip) speech metrics show statistically significant differences between controls, bulbar symptomatic and bulbar pre-symptomatic patients. We report on the sensitivity and specificity of these metrics using five-fold cross-validation.  We further conducted a LASSO-LARS regression analysis to uncover the relative contributions of various acoustic and visual features in predicting the severity of patients' ALS (as measured by their self-reported ALSFRS-R scores). Our results provide encouraging evidence of the utility of automatically extracted audiovisual analytics for scalable remote patient assessment and monitoring in ALS. 
\end{abstract}

\section{Multimodal Conversational Agents for Health Monitoring}
The development of technologies that rapidly diagnose medical conditions, recognize pathological behaviors, continuously monitor patient status, and deliver just-in-time interventions using the user's native technology environment remains a critical need today~\cite{kumar2012mobile}. The COVID-19 pandemic has further highlighted the need to make telemedicine and remote monitoring more readily available to patients with chronic neurological disorders \cite{bombaci2020telemedicine}. However, early detection or progress monitoring of neurological or mental health conditions, such as clinical depression, ALS, Alzheimer’s disease, dementia, etc., is often challenging for patients due to various reasons, including, but not limited to: (i) no access to neurologists or psychiatrists; (ii) lack of awareness of a given condition and the need to see a specialist; (iii) lack of an effective standardized diagnostic or endpoint; (iv) substantial cost and transportation involved in conventional or traditional solutions; and in some cases (v) lack of medical specialists in these fields~\cite{steven2013can}.\\
The NEurological and Mental health Screening Instrument (NEMSI)~\cite{suendermann2019nemsi}, has been developed to bridge this gap. NEMSI is a cloud-based multimodal dialog system that can be used to elicit evidence required for detection or progress monitoring of neurological or mental health conditions through automated screening interviews conducted over the phone or via web browser.
While intelligent virtual agents have been proposed in previous work for such diagnosis and monitoring purposes~\cite{devault2014a,lisetti2015a}, NEMSI offers three significant innovations: First, NEMSI uses readily available devices (web browser or mobile app), in contrast to dedicated, locally administered hardware, like cameras, servers, audio devices, etc. Second, NEMSI’s backend is deployed in an automatically scalable cloud environment allowing it to serve an arbitrary number of end-users at a small cost per interaction. Thirdly, the NEMSI system is natively equipped with real-time analytics modules that extract a variety of speech and video features of direct relevance to clinicians in the neurological space, such as speech and pause duration for the assessment of ALS, or geometric features derived from facial landmarks for the automated detection of orofacial impairment in stroke.\\
This paper investigates the utility of audio and video metrics collected via  NEMSI for early diagnosis and monitoring of ALS. We specifically investigate two research questions. First, which metrics demonstrate statistically significant differences between (a) healthy controls and bulbar pre-symptomatic people with ALS or pALS (thereby assisting in early diagnosis), as well as (b) bulbar-presymptomatic patients and bulbar symptomatic patients (thereby assisting in progress monitoring)? Second, for pALS cohorts, which metrics are most predictive of their self-reported ALS Functional Rating Scale-Revised (ALSFRS-R~\cite{cedarbaum1999alsfrs}) score? Before addressing these questions, however, we briefly summarize the current state of ALS research, and describe our data collection and metrics extraction process. 

\begin{table*}[th]
    \centering
    \begin{tabular}{lccccc}
        \hline \hline
         & \multicolumn{2}{c}{Sex} & & ALSFRS-R & ALSFRS-R \\
        Group & Female & Male & Age (years) & Total & Bulbar  \\ \hline \hline
        CON & 68 & 14 & 43; 41.62 (19.00) & 48; 47.89 (0.94) & 12; 11.94 (0.36) \\
        BUL & 17 & 15 & 63; 59.56 (10.29) & 36; 33.09 (7.46) & 9; 8.75 (1.57) \\
        PRE & 12 & 10 & 61; 57.18 (11.31) & 40; 36.45 (8.58) & 12; 12.00 (0.00) \\
        \hline \hline
    \end{tabular}
    \caption{Participant characteristics for the three groups -- controls (CON), bulbar symptomatic (BUL), and bulbar pre-symptomatic (PRE). Age and ALSFRS-R scores are presented as: median; mean (standard deviation).}
    \label{tab:participant_stats}
    \vspace{-6mm}
\end{table*}

\section{Current State of ALS Diagnosis and Monitoring}
\label{sec:sota}

ALS is a neurodegenerative disease that affects roughly 4 to 6 people per 100,000 of the general population \cite{majoor2003genetic,al2013epidemiology}. Early detection and continuous monitoring of ALS symptoms is crucial to provide optimal patient care \cite{paganoni2014diagnostic}. For instance, decline in speech intelligibility is said to have a negative impact on patients' quality of life \cite{chio2004cross,joubert2011speech}, and continuous monitoring of speech intelligibility could prove valuable in terms  of patient care. Traditionally, subjective measures, such as patient-reports or ratings by clinicians, are used to detect and monitor speech impairment. However, recent studies show that objective measures allow for earlier detection of ALS symptoms \cite{yunusova2013speech,allison2017a,gomez2017articulation,norel2018detection,bandini2018kinematic,perry2018lingual,wisler2021}, stratification and classification of patients \cite{rong2020speech} and can provide markers for disease onset, progression and severity  \cite{rong2016predicting,wang2018automatic,makkonen2018speech,wilson2019detecting,barnett2020reliability}. 
These objective measures can be automatically extracted, thereby allowing for more frequent monitoring, potentially improving treatment. 
The success of the Beiwe Research Platform\footnote{\texttt{https://www.beiwe.org/}}~\cite{berry2019a} to track ALS disease progression demonstrates the viability of such remote monitoring solutions.  

\section{Methods}
\label{sec:data}

\subsection{Collection Setup}
\label{sec:system}


NEMSI end users are provided with a website link to the secure screening portal and login credentials by their caregiver or study liaison (physician, clinic, a referring website or patient portal). After completing 
microphone and camera checks, subjects participate in a conversation with ``Nina'', a virtual dialog agent. Nina's virtual image appears in a web window, and subjects are able to see their own video. 
During the conversation, Nina engages subjects in a mixture of structured speaking tasks and open-ended questions to elicit speech and facial behaviors relevant for the type of condition being screened for.

Analytics modules automatically extract speech (e.g., speaking rate, duration measures, fundamental frequency (F0)) and video features (e.g., range and speed of movement of various facial landmarks) in real time and store them in a database, along with meta-data about the interaction, such as call duration and completion status. All this information can be accessed by the study liaison  
through an easy-to-use dashboard, which provides a 
summary of the interaction (including access to a video recording and the analytic measures computed), as well as a detailed breakdown of the metrics by individual interaction turns.

\subsection{Data}
\label{sec:data}
Data from 136 participants (see Table~\ref{tab:participant_stats}) were collected between September 2020 and March 2021 in cooperation with EverythingALS and the Peter Cohen Foundation\footnote{\url{https://www.everythingals.org/research}}. 
For this cross-sectional study we included one dialog session per subject.\footnote{If a subject participated in multiple dialog sessions, we took the first successful one, i.e., the first \textit{complete} call for which valid  
metrics were extracted and all ALSFRS-R questions were answered.}


The conversational protocol elicits five different types of speech samples from participants, inspired by prior work \cite{silbergleit1997acoustic, tomik2010dysarthria,novotny2020comparison,agurto2019}: (a) sustained vowel phonation, (b) read speech, (c) measure of diadochokinetic rate (rapidly repeating the syllables /p\textscripta t\textscripta k\textscripta/), and (d) free speech (picture description task). For (b) read speech, the dialog contains six speech intelligibility test (SIT) sentences of increasing length (5 to 15 words), and one passage reading task (Bamboo Passage; 99 words).
After dialog completion, participants filled out the \emph{ALS Functional Rating Scale-revised  (ALSFRS-R)}, a standard instrument for monitoring the progression of ALS~\cite{cedarbaum1999alsfrs}.
The questionnaire consists of 12 questions about physical functions in activities of daily living. Each question provides five answer options, ranging from \emph{normal function} (score 4) to \emph{severe disability} (score 0). The total ALSFRS-R score is the sum of all sub-scores (therefore ranging from 0 to 48).
The ALSFRS-R comprises four scales for different domains affected by the disease: bulbar system, fine and gross motor skills, and respiratory function. 

We stratified subjects into three groups for statistical analysis: (a) Healthy controls (\textit{CON}); 
(b) pALS with a bulbar sub-score $<12$ (first three ALSFRS-R questions) were labeled bulbar symptomatic (\textit{BUL}); 
and (c) pALS with a bulbar sub-score of 12 were labeled bulbar pre-symptomatic (\textit{PRE}). 
Similar to \cite{allison2017a} we aim at identifying acoustic and visual speech measures that show significant differences between these groups. 



\section{Signal Processing and Metrics Extraction}
\label{sec:metrics}

\subsection{Acoustic Metrics}
We use measures commonly established for clinical speech analysis with regard to ALS~\cite{allison2017a}, including \textit{timing measures}, 
\textit{frequency domain measures}, 
and measures specific to the diadochokinesia task (DDK), such as syllable rate and cycle-to-cycle temporal variation~\cite{rong2020automated}.
Table~\ref{tab:speech} shows the metrics and speech task types from which they are extracted. Additionally, speech intensity (mean energy in dB SPL excluding pauses) was extracted for all utterances.
The picture description task (free speech) was not used for this analysis.

All acoustic measures were automatically extracted with the speech analysis software Praat~\cite{boersma2001speak}.
Speaking and articulation rates 
are computed based on \textit{expected} number of words because forced alignment is error-prone for dysarthric speech \cite{yeung2015improving}. For that reason, these measures can be noisy, if for example a patient did not finish the reading passage. Hence, we automatically remove outliers based on thresholds for the Bamboo task: speaking rates$>250$ words/min, articulation rates$>350$ words/min, and PPT$>80\%$ are excluded.\footnote{The outlier thresholds and the tasks for which they are applied were determined by manual inspection of the data.}


\begin{table}[]
    \centering
    \resizebox{\linewidth}{!}{
    \begin{tabular}{r|l}
    \hline \hline
    Speech type & Collected metrics \\ \hline \hline
    Held vowel & Mean F0 (Hz), jitter (\%), shimmer (\%), \\
     & HNR (dB), CPP (dB) \\ \hline
    SIT and & Speaking and articulation duration (sec), \\
    Bamboo & speaking and articulation rate (words/min), PPT \\ 
    \hline
    DDK & Speaking and articulation duration (sec), \\
     & Syllable rate (syllables/sec), number of \\
     & produced syllables, cTV (sec) \\
     \hline \hline
    \end{tabular}
    }
    \caption{Extracted acoustic metrics. PPT~= percent pause time, DDK~= diadochokinesia, CPP = cepstral peak prominence, HNR = harmonics-to-noise ratio, cTV~= cycle-to-cycle temporal variation.}
    \label{tab:speech}
    \vspace{-4mm}
\end{table}


\subsection{Visual Metrics}\label{subsec:visualmetrics}

Facial metrics were calculated for each utterance in three steps: (i)~face detection using the \textit{Dlib}\footnote{http://dlib.net/} face detector, which uses five histograms of oriented gradients to determine the (x, y)-coordinates of one or more faces for every input frame~\cite{dalal2005a}, (ii)~facial landmark extraction using the Dlib facial landmark detector, which uses an ensemble of regression trees proposed in~\cite{kazemi2014a} to extract 68 facial landmarks according to MultiPIE~\cite{gross2010a}, and (iii)~facial metrics calculation, which uses 20 facial landmarks to compute the metrics shown in Table~\ref{tab:facial_metrics} (cf. \cite{neumann2020utility} for details). 
Finally, all facial metrics in pixels were normalized within every subject by dividing the values by the inter-lachrymal distance in pixels (measured as distance between the right corner of the left eye and the left corner of the right eye) for each subject. 

\begin{table}[]
    \centering
    \begin{tabular}{r|l|l}
        \hline \hline
        Category & Metrics & Description \\ \hline \hline
        Move- & open & Lips opening and \\
        ment & width & mouth width; \\
         & LL\_path & Displacement of lower \\
         & JC\_path & lip and jaw center; \\
         & eye\_open & Eye opening; \\
         & eyebrow\_vpos & Vertical eyebrow \\
         &  & displacement \\
        \hline
        Velocity & vLL & Velocity and \\
         & vJC & speed of lower  \\
         & vLL\_abs  & lip and jaw center \\
         & vJC\_abs & (px/frames) \\
        \hline
        Accel- & aLL & Acceleration of lower \\
        eration & aJC & lip and jaw center \\
         & aLL\_abs  & (px/frames$^2$) \\
         & aJC\_abs &  \\
        \hline
        Jerk & jLL & Jerk of lower \\
         & jJC &  lip and jaw center \\
         & jLL\_abs  & (px/frames$^3$) \\
         & jJC\_abs &\\
        \hline
        Surface & S (S\_R, S\_L) & Area of the mouth \\
         & & (right and left half, px$^2$) \\
         & S\_ratio\_avg & Mean symmetry ratio \\
        \hline
        Eye blinks & eye\_blinks & Eye blinks per sec. \\
        \hline \hline
    \end{tabular}
    \caption{Extracted facial metrics. Maximum (suffix \_max) and average (\_avg) were extracted for movement and surface metrics, and maximum, average and minimum (\_min) were extracted for velocity, acceleration and jerk metrics.}
    \label{tab:facial_metrics}
    \vspace{-4mm}
\end{table}

\section{Analyses and Observations}
\label{sec:analysis}

To normalize for sex-specific differences in metrics (such as F0), we z-scored all metrics by sex group. Additionally, all metrics reported below (except speaking and articulation duration) were averaged across speech task type. 
An important caveat to all the analyses presented here is the imbalance of sample size between the cohorts; also, future extensions to this work will need a larger sample size of the BUL and PRE cohorts to make robust and generalizable statistical claims.  

\definecolor{af}{rgb}{0.6470588235294118, 0.16470588235294117, 0.16470588235294117}
\definecolor{ff}{rgb}{0.0, 0.5450980392156862, 0.5450980392156862}
\begin{figure}[h]
    \centering
    \includegraphics[trim=10 12 10 10, clip, width=\linewidth]{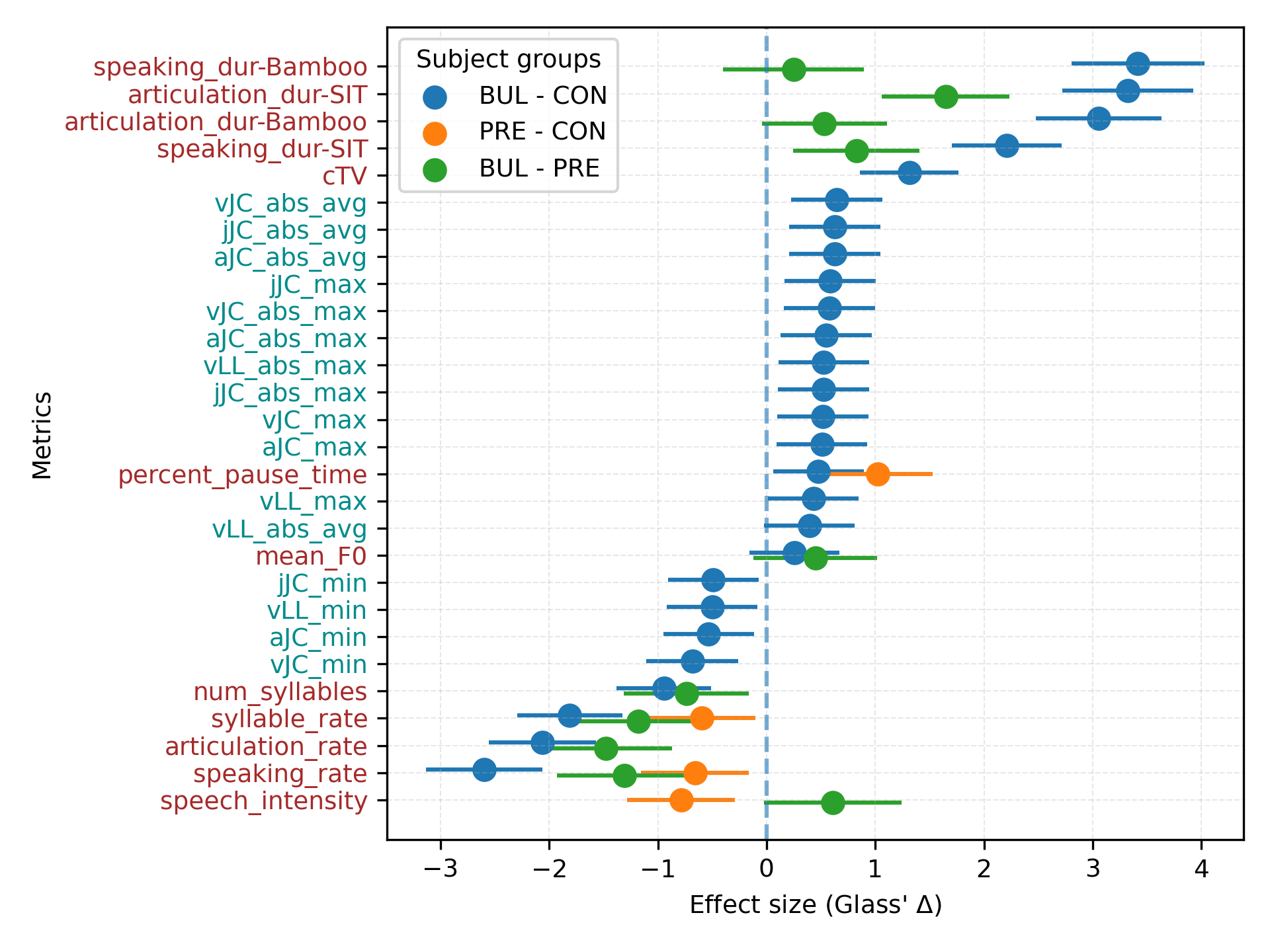}
    \caption{Effect sizes of \textbf{\textcolor{af}{acoustic}} and \textbf{\textcolor{ff}{visual}} metrics that show statistically significant differences at $p<0.05$. Effect sizes are shown with a 95\% confidence interval and are ranked by the BUL--CON group pair.}
    \label{fig:effect_sizes1}
    \vspace{-5mm}
\end{figure}

\begin{figure*}
    \vspace{-10mm}
     \centering
     \begin{subfigure}[b]{0.33\textwidth}
         \centering
         \includegraphics[trim=20 5 20 20, clip, width=\textwidth]{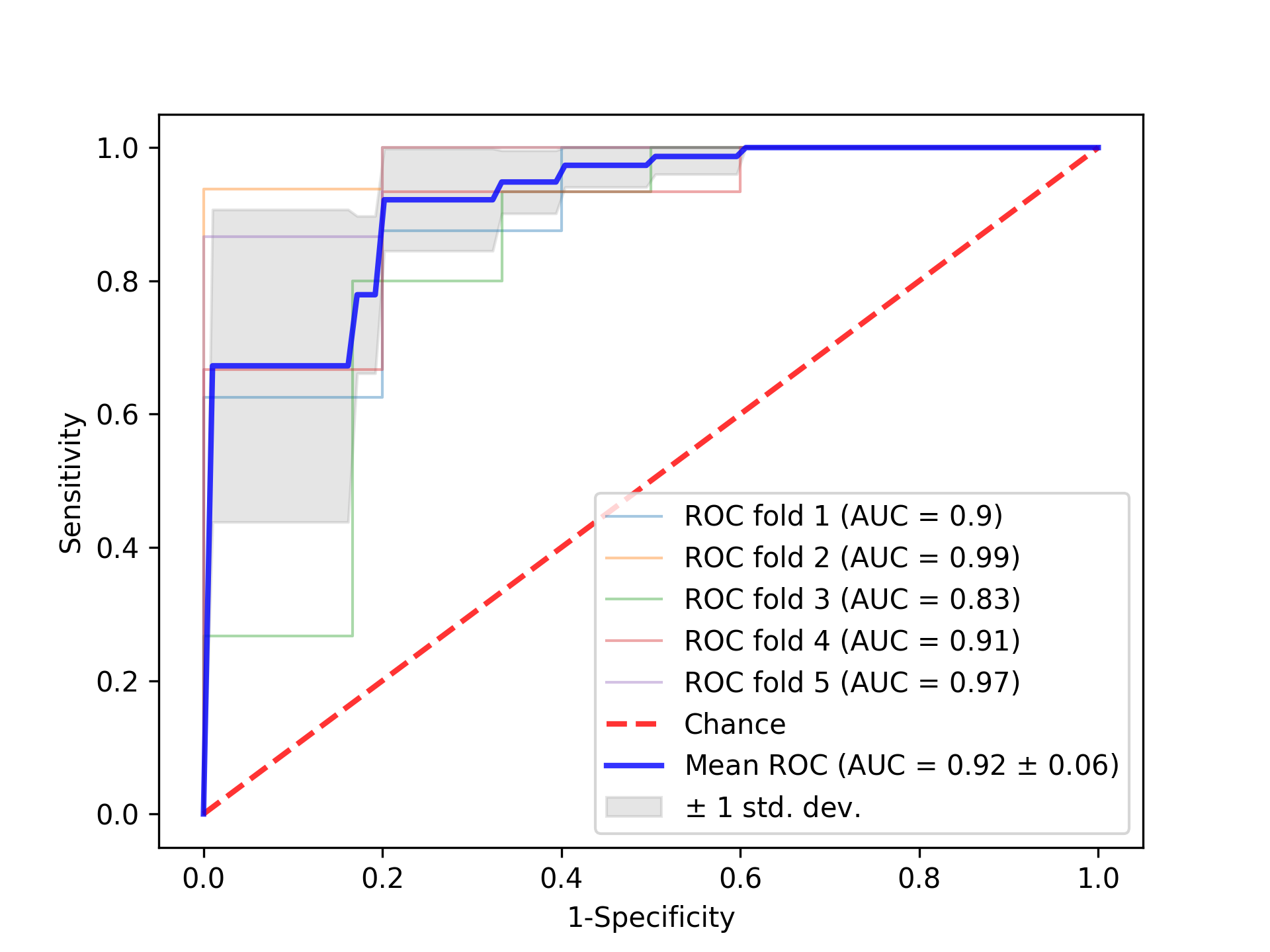}
         \caption{BUL vs CON}
         \label{fig:roc_bul}
     \end{subfigure}
     \hfill
     \begin{subfigure}[b]{0.33\textwidth}
         \centering
         \includegraphics[trim=20 5 20 20, clip, width=\textwidth]{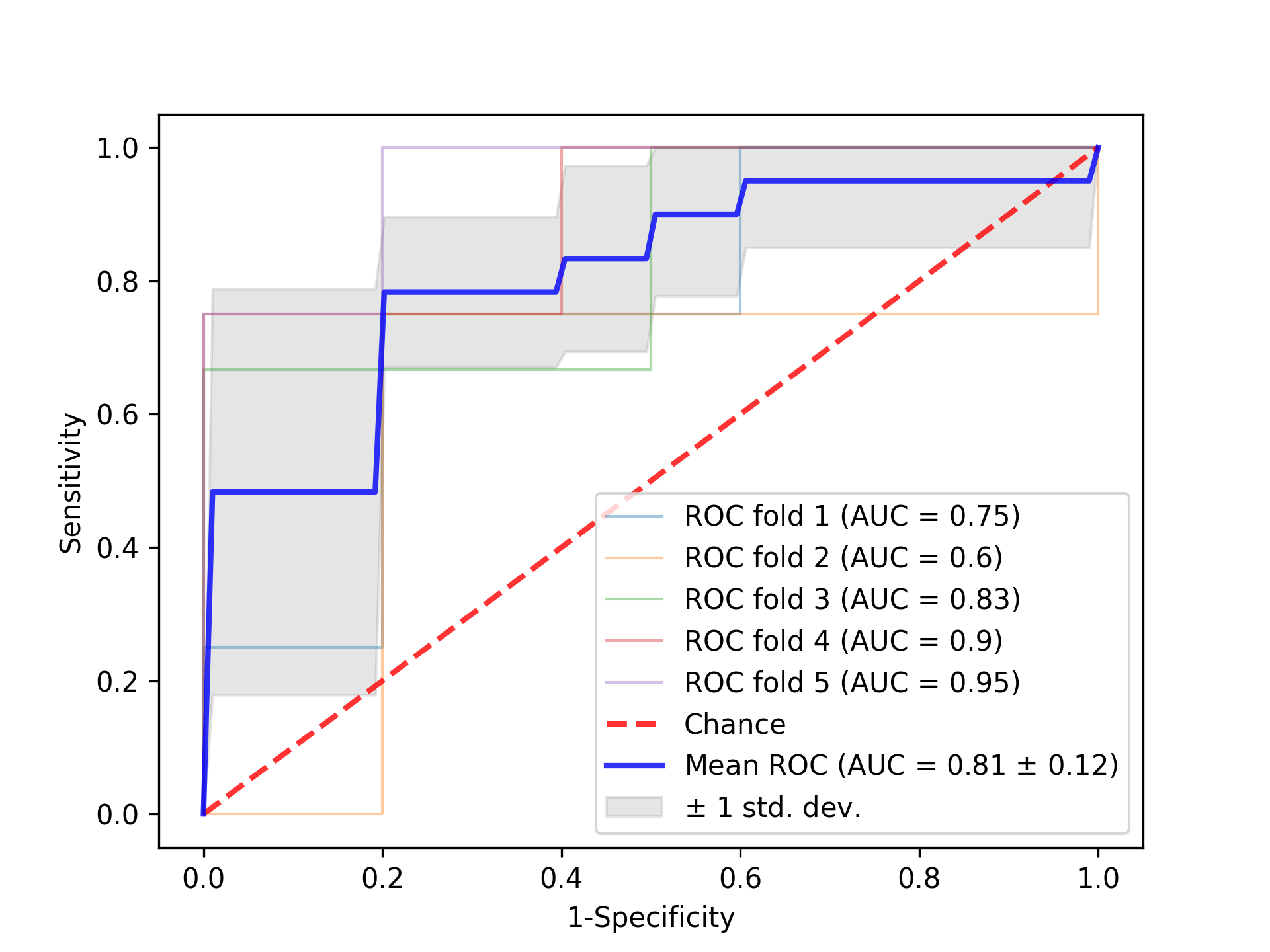}
         \caption{BUL vs PRE}
         \label{fig:roc_pre}
     \end{subfigure}
     \hfill
     \begin{subfigure}[b]{0.33\textwidth}
         \centering
         \includegraphics[trim=20 5 20 20, clip, width=\textwidth]{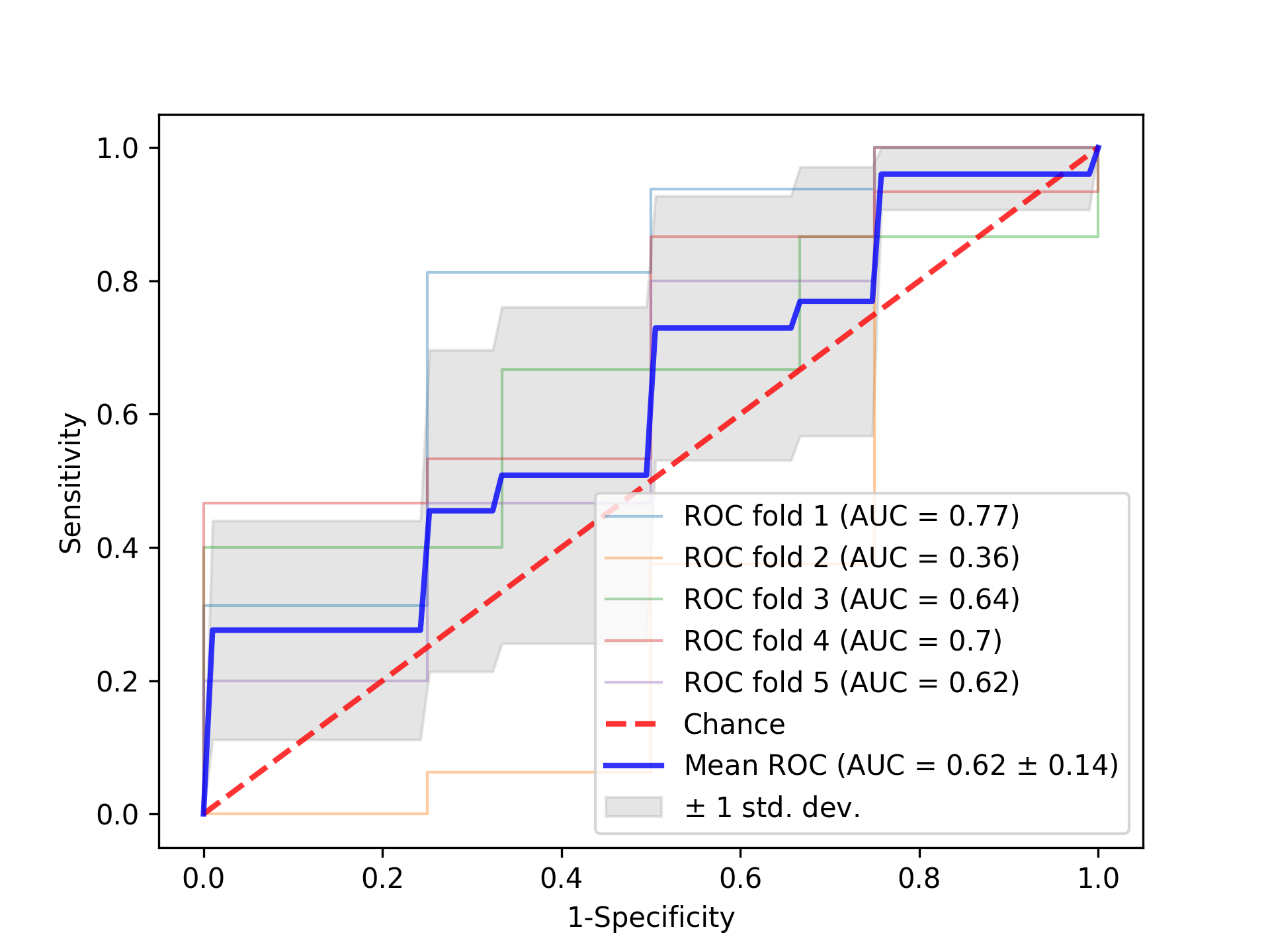}
         \caption{PRE vs CON}
         \label{fig:roc_con}
     \end{subfigure}
        \caption{ROC curves displaying the performance of binary classification with 5-fold crossvalidation for all group pairs.}
        
        \label{fig:roc}

\end{figure*}

\begin{figure*}
    \centering
    \begin{subfigure}{.45\textwidth}
    \centering
    \includegraphics[trim=15 10 10 10, clip, width=\textwidth]{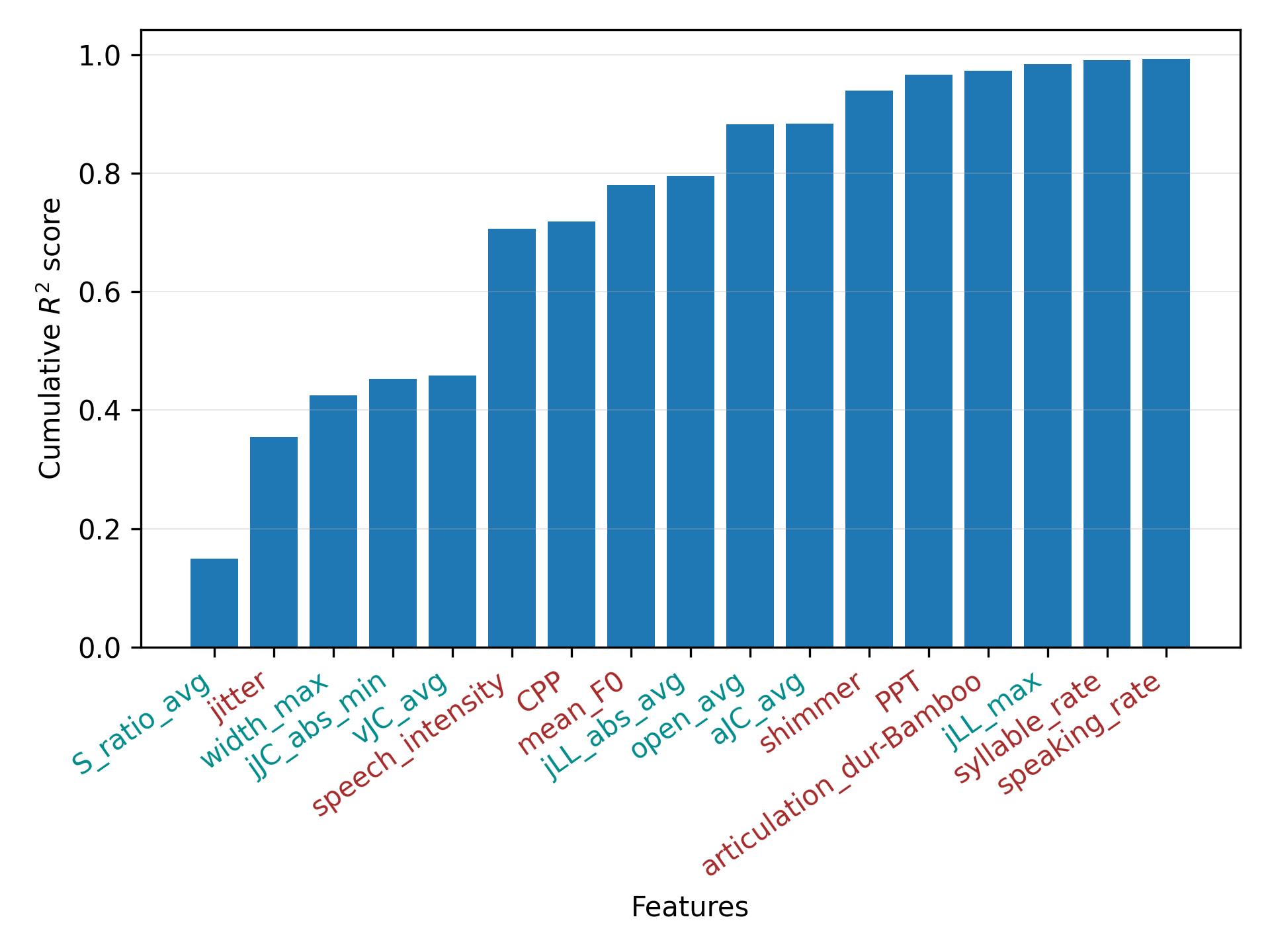}
    \caption{Final 17 features obtained for PRE group} 
    \vspace{-2mm}
    \label{fig:LASSO_PRE}
    \end{subfigure}
    \begin{subfigure}{.45\textwidth}
    \centering
    \includegraphics[trim=15 10 10 10, clip,  width=\textwidth]{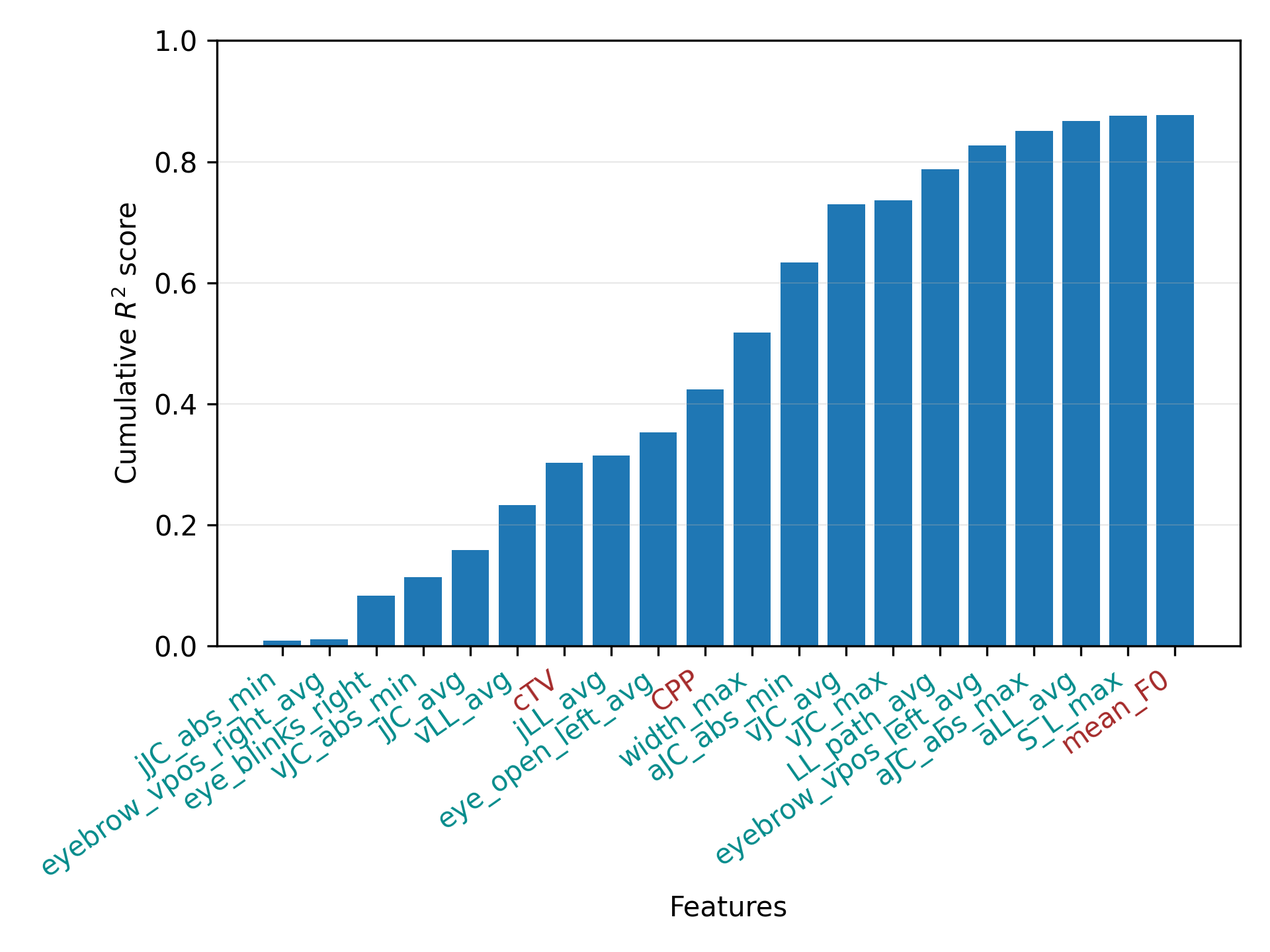}
    \caption{First 20 features (out of 23) for BUL group} 
    \vspace{-2mm}
    \label{fig:LASSO_BUL}
    \end{subfigure}
    \caption{\textbf{\textcolor{af}{Acoustic}} and \textbf{\textcolor{ff}{visual}} features from the LASSO LARS regression path.}
    \vspace{-4mm}
    \label{fig:lasso_lars}
\end{figure*}

\subsection{RQ1: Which metrics demonstrated statistically significant pairwise differences between controls, bulbar pre-symptomatic and bulbar symptomatic pALS cohorts?}

We conducted a non-parametric Kruskal-Wallis test for every acoustic and facial metric to identify the metrics that showed a statistically significant difference between the cohorts. For all metrics with $p<0.05$ a post-hoc analysis was done (again Kruskal-Wallis) between every combination of two cohorts to find out which groups can be distinguished. 
Figure~\ref{fig:effect_sizes1} shows effect size, measured as Glass' $\Delta$ \cite{glass1981meta}, for all metrics that show statistically significant difference ($p<0.05$) between different subject groups.


In addition to the statistical tests, we conducted 5-fold cross-validation with logistic regression to 
investigate binary classification performances, and in turn sensitivities and specificities, of our aforementioned metrics in distinguishing the CON vs PRE (with applications to early diagnosis) and PRE vs BUL (progress monitoring) groups. We investigated using both the full feature set as well as feature selection with recursive feature elimination for classification and found that the latter performed better as the former method tends to overfit the data, given our small sample size. 
Receiver operating characteristics (ROC) curves for these classification experiments encapsulating sensitivities and specificities are presented in Figure~\ref{fig:roc}. 
For the CON vs PRE case, we observed that the mean unweighted average recall (UAR) across 5 cross-validation folds was 0.63 $\pm$ 0.08, respectively, significantly above chance, suggesting promising applications for early diagnosis. For the PRE vs BUL case (and therefore progress monitoring), these numbers were even better -- 0.77 $\pm$ 0.05, respectively. The BUL vs CON case, unsurprisingly, produced the best results. 

Looking at acoustic features, we found that timing measures (speaking and articulation duration and rate; PPT; syllable rate, cTV) exhibit strong differences between groups and that the effect sizes of these metrics are highest between the BUL group and the CON group. 
Mean F0 also showed a significant difference with small effect sizes. 
%
%
For visual metrics, the results indicate that velocity, acceleration and jerk measures are generally the best indicators for ALS. Additionally, while the jaw center (JC) seems to be more important than the lower lip (LL) for detecting ALS, further investigations are necessary to ensure that the difference between the JC and LL metrics is not just due to a difference in facial landmark detection accuracy. 


\subsection{RQ2: Which metrics contribute the most toward predicting the ALSFRS-R score?}

In a regression analysis, we investigated the predictive power of the extracted metrics with regard to both BUL and PRE pALS cohorts. 
To investigate this, we employed a LASSO (least absolute shrinkage and selection operator) regression with the objective to predict the total ALSFRS-R score (implemented using least-angle regression (LARS) algorithm~\cite{efron2004least}). The algorithm is similar to forward stepwise regression, but instead of including features at each step, the estimated coefficients are increased in a direction equiangular to each one’s correlations with the residual.




 

Figure~\ref{fig:LASSO_PRE} shows the final 17 features, in the order they were selected by the LASSO-LARS regression, on data from 19 PRE samples\footnote{The number of samples in the classification and regression analyses differ from Table~\ref{tab:participant_stats} because not all metrics were present for all subjects, either because the task was not performed correctly or system errors.}  
along with the cumulative model $R^2$ at each step. We observe that both facial metrics (mouth opening and symmetry ratio, higher order statistics of jaw and lips) and acoustic metrics (particularly voice quality metrics such as jitter, shimmer and mean F0) added useful predictive power to the model, 
suggesting that these might be useful in modeling severity in bulbar pre-symptomatic pALS. 
 
On the other hand, for the BUL cohort, Figure~\ref{fig:LASSO_BUL} shows a slightly different set of 20 features obtained using LASSO-LARS (based on 26 participants). We observe that facial metrics (eye blinks and brow positions, in addition to higher order statistics of jaw and lips) add more predictive power to the model than acoustic metrics (such as cTV, CPP and mean F0), 
suggesting that these might find utility in modeling severity in bulbar symptomatic pALS. 


\section{Conclusions}

Our findings demonstrate the utility of multimodal dialog technology for assisting early diagnosis and monitoring of pALS. Multiple automatically extracted acoustic (rate, duration, voicing) and visual (higher order statistics of the jaw and lip) speech metrics show significant promise in assisting with both early diagnosis of bulbar pre-symptomatic ALS vs healthy controls, as well as for progress monitoring in pALS. Moreover, using LASSO-LARS to model the relative contribution of these features in predicting the ALSFRS-R score highlights the utility of incorporating different speech and facial metrics for modeling severity in bulbar pre-symptomatic and bulbar symptomatic pALS. While higher order statistics of the jaw and lower lip facial features and timing, pausing and rate-based speech features were useful across the board for both cases, voice quality and mouth opening and area metrics seem to be more useful for the bulbar pre-symptomatic group, while spectral and eye-related metrics are relevant for the bulbar symptomatic group. Future work will expand these analyses to more speakers to ensure the statistical robustness and generalizability of these trends. 

\bibliographystyle{IEEEtran}
\bibliography{mybib}

\end{document}